\documentstyle[12pt,epsfig]{article}
\newcommand{\beq}{\begin{equation}}
\newcommand{\eeq}{\end{equation}}
\newcommand{\bea}{\begin{eqnarray}}
\newcommand{\eea}{\end{eqnarray}}

\newcommand{\gsim}{\lower.7ex\hbox{$
\;\stackrel{\textstyle>}{\sim}\;$}}
\newcommand{\lsim}{\lower.7ex\hbox{$
\;\stackrel{\textstyle<}{\sim}\;$}}

\def\lsim{\mathrel{\rlap{\lower3pt\hbox{\hskip0pt$\sim$}}
    \raise1pt\hbox{$<$}}}         
\def\gsim{\mathrel{\rlap{\lower4pt\hbox{\hskip1pt$\sim$}}
    \raise1pt\hbox{$>$}}}         

\renewcommand{\Im}{{\rm Im}\,}

\newcommand{\bibit}[1]{\bibitem{#1}}

\newcommand{\aver}[1]{\langle #1\rangle}

\newcommand{\La}{\overline{\Lambda}}
\newcommand{\Lam}{\Lambda_{\rm QCD}}

\newcommand{\as}{\alpha_s}
\newcommand{\GeV}{\,\mbox{GeV}}
\newcommand{\MeV}{\,\mbox{MeV}}
\newcommand{\matel}[3]{\langle #1|#2|#3\rangle}
\newcommand{\state}[1]{|#1\rangle}

\newcommand{\vep}{\varepsilon}

\begin{document}

\begin{titlepage}
\renewcommand{\thefootnote}{\fnsymbol{footnote}}

\begin{flushright}
Bicocca-FT-01/24\\
UND-HEP-01-BIG\hspace*{.08em}07\\
hep-ph/0111166\\
\end{flushright}
\vspace*{1.3cm}

\begin{center} \Large

{\bf On the chromomagnetic expectation value $\mu_G^2$ 
and higher power corrections \\
in heavy flavor mesons}
\end{center}
\vspace*{5mm}
\begin{center} {\Large\tt
Nikolai Uraltsev} \\
\vspace{1.2cm}
{\normalsize
{\it INFN, Sezione di Milano, Milan, Italy$^{\:*}$\\
{\small \rm and} \\
{\it ~\hspace*{-15mm} Department of Physics, University of Notre 
Dame du Lac, Notre Dame, IN 46556, U.S.A.$^*$ \hspace*{-15mm}~}
}\\
}

\normalsize
\vspace*{25mm}

{\large{\bf Abstract}}\vspace*{-1mm}\\
\end{center}
The important parameter $\mu_G^2$ of the heavy quark expansion is
analyzed including perturbative and power corrections. It is found that
$\mu_G^2(2\GeV)$ is known with a few percent accuracy. The
perturbative corrections are computed and found small. A nonperturbative
relation is suggested which allows to control the power corrections. 
We conclude that $\mu_G^2(1\GeV)\!=\!0.35 \,^{+ 0.03}_{-0.02}\,\GeV^2$.
The two-loop expression for the 
effective ``\hspace*{-1.5pt}{\it ME}\,'' radiation 
coupling $\as^{(me)}(\omega)$ is given which improves reliability of the 
perturbative evolution of $\mu_G^2$ towards the low momentum scale. On 
the nonperturbative side,
we advocate the utility of combining the heavy quark expansion with
expanding around the ``BPS''-type approximation for the meson
wavefunction, which implies 
relations $\mu_\pi^2\!\approx \!\mu_G^2$ and 
$-\rho_{LS}^3\!\approx\!\rho_D^3$ as well as similar ones for the 
nonlocal correlators.

\vfill

~\hspace*{-7mm}\hrulefill \hspace*{3cm} \\
\footnotesize{\noindent $^*$On leave of absence from 
St.\,Petersburg Nuclear Physics Institute, Gatchina, St.\,Petersburg 
188300, Russia}

\noindent
\end{titlepage}

\newpage

The heavy quark expansion allows to quantify the effects of nonperturbative
physics in beauty decays, often in a model-independent way starting
from the first principles of QCD. The most informative predictions are
obtained for observables where the Operator Product Expansion (OPE)
applies, like the inclusive decay widths. The leading nonperturbative
effects are described by the two heavy quark expectation values
$\mu_\pi^2$ and $\mu_G^2$ of the kinetic and chromomagnetic operators,
respectively, entering at the $\Lam^2/m_b^2$
level \cite{buvbs}. It turns out, in particular, that the extractions
of $V_{cb}$ are sensitive to them.

The value of $\mu_G^2$ can be determined from the mass
splitting between $B^*$ and $B$ mesons. The kinetic expectation value
$\mu_\pi^2$ is {\it a priori\,} more uncertain. The set of heavy 
quark sum rules,
however, significantly restricts $\mu_\pi^2$ in terms of $\mu_G^2$
\cite{rev,ioffe}. In particular, $\mu_\pi^2 \!-\! \mu_G^2$ is positive, yet
unlikely to exceed $0.2\GeV^2$. 

In this paper we analyze the value of $\mu_G^2$. Throughout the paper it 
it refers to the infinite mass limit. In QCD the heavy
quark operators depend on the renormalization point. We consistently
take this into account using the complete definition of these operators
in the framework of the quantum field theory suggested earlier
\cite{fiveblmope}. To this end the perturbative correction to the
hyperfine splitting is computed. 

We also address the power corrections. A nonperturbative relation is
suggested which looks well satisfied in QCD. It allows one to
control a number of higher-order effects. We conclude that the
associated uncertainty in $\mu_G^2$ is rather small. 

We propose and briefly discuss another theoretical tool based on the 
smallness of the difference $\mu_\pi^2\!-\!\mu_G^2$, which can shed
light on the structure of the higher-order power corrections and improve
the accuracy of certain applications of the heavy quark expansion to
charm.

\section{Perturbative effects}

The value of $\mu_G^2(\mu)$ is extracted in practice using the relation
$\mu_G^2=\frac{3}{2} m_b (M_{B^*}\!-\!M_B)$. There are perturbative
corrections to this relation depending logarithmically on $\mu/m_b$. At
finite $m_b$ there are also nonperturbative corrections suppressed by 
powers of $1/m_b$. 

The bare heavy quark operator $O_G=\bar{Q}
\frac{i}{2}g_s\sigma_{\mu\nu}G^{\mu\nu} Q = -\bar{Q} g_s \vec{\sigma}
\vec{B} Q$ is ultraviolet divergent in the non-Abelian theory. The
normalization point $\mu$ is introduced via the upper cutoff in the
integral over the antisymmetric small velocity (SV) heavy quark 
structure function $W_-(\vep)$
expressing the sum rule for the matrix elements of $\mu_G^2$. In the
standard heavy quark notation it takes the following form:
\beq
\frac{\mu_G^2(\mu)}{3} = \int_0^\mu W_-(\vep)\, \vep^2
{\rm d}\vep = 2\sum_{\vep_m< \mu} \vep_m^2 |\tau_{3/2}|^2 - 2\sum_{\vep_n<
\mu} \vep_n^2 |\tau_{1/2}|^2\,,
\label{10}
\eeq
where $\tau$'s are the $P$-wave transition amplitudes and $\vep_k$ are
the corresponding excitation energies (for a review, see \cite{ioffe}).
The OPE gives the SV structure functions in terms of the zero-recoil 
matrix elements
of the momentum operators $\pi_j \!=\! \bar{Q} iD_j Q$: 
$$
\vep^2 W(\vep) \propto \sum_n \matel{H_Q}{\pi_j}{n_{\vec{p}}}\, 
\matel{n_{\vec{p}}}{\pi_l}{H_Q}\, \delta^3(\vec{p}\,) \,
\delta(E_n-\vep)\;.
$$
Since $[D_j, D_l]= \!-ig_s G_{jl} = i \epsilon_{jlk} g_s B_k$ holds, 
the sum rule
(\ref{10}) is transparent \cite{rev}: $ \bar{Q} \vec \pi^{\,2} Q =
\sum_k \pi_k \pi_k\,$, and  $ \,\bar{Q} g_s B_l Q 
= i\epsilon_{ljk}\pi_j \pi_k$. 

In QED the integral in Eq.~(\ref{10}) converges and defines the magnetic
field strength $e\vec{B}_{{\rm em}}(0)$ at the position of the static
center. The magnetic spin interaction of an elementary heavy fermion is
given precisely by this expectation value times the Dirac anomalous
moment, $e\vec{B}_{{\rm em}}\vec\sigma \,\frac{1}{2m}\!
\left(1+\frac{\alpha}{2\pi} + ...\right)$. In a non-Abelian theory like
QCD the integral diverges in the ultraviolet, and the expectation value
of $g_s\vec{B}_{{\rm chr}}(0)$ depends on the normalization point. Let us
note that the adopted definition of the operator corresponds to the
usual scheme with the two covariant derivatives taken at different
points and connected by the $P$-exponent. The displacement 
lies on the (Euclidean) time axis and its magnitude is governed by
$1/\mu$. More precisely,
\beq
\aver{\bar{Q} \pi_j (x_0) P\, {\rm e}^{\,i\!\int_{0}^{x_0} A_0(x) {\rm d}
x_0} 
\pi_k Q(0)}
= \int_0^\infty x_0 \,{\rm e}^{-\mu x_0} {\rm d} \mu \;
\aver{\bar{Q} \pi_j \pi_k Q(0)}_\mu\;.
\label{14}
\eeq

The heavy quark Hamiltonian has the well known form
\beq
{\cal H}_Q = m_Q - g_s A_0 + \frac{\vec{\pi}^{\,2}}{2m_Q} + 
c_G \frac{-\frac{i}{2}g_s\sigma_{jk}G_{jk}}{2m_Q} + 
{\cal O}\left(\frac{1}{m_Q^2}\right)
\;.
\label{16}
\eeq
To compute perturbatively the Wilson coefficient $c_G(\mu)$ we therefore
can consider the zero-velocity heavy quark transition amplitude 
$T_{ij}(\omega)$
mediated by the currents $\bar{Q} iD_i Q$ and $\bar{Q} iD_j Q$, Fig.~1
(the corresponding OPE formalism is discussed in detail, e.g.\ in
Ref.~\cite{ioffe}). To select the matrix element of the chromomagnetic 
operator we 
evaluate it on a heavy quark state including scattering of an
additional (transverse) gluon, and look for the component antisymmetric in
$i,j$. This appears in the linear in the gluon momentum
$\vec{q}$ approximation. For simplicity, we take $q_0\!=\!0$, and assume the
initial quark is at rest. We also do not show explicitly the heavy quark
spin indices (as if $Q$ is a scalar). 

\begin{figure}[hhh]
 \begin{center} \vspace*{-6mm}
 \mbox{\epsfig{file=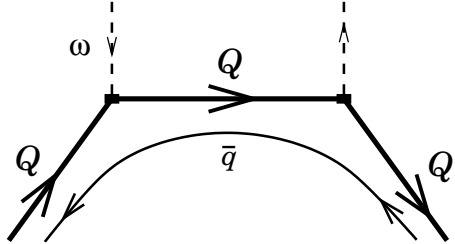,width=6.0cm}} \vspace*{-6mm}
  \end{center}
\caption{
{\small
The heavy quark forward scattering amplitude as a function of energy
$\omega$. The solid blocks denote the momentum operator 
$\bar{Q} i\vec{D} Q$.}
}\vspace*{-2mm}
\end{figure}

The tree level ${\cal O}(\alpha_s^0)$ expression is obvious:
\beq
T_{ij}(\omega) = \frac{1}{-\omega + i\epsilon} \, g_s\, q_j \delta_{il}
\, \bar{Q} \mbox{$\frac{\lambda^c}{2}$} Q
\label{18}
\eeq
where $l$ and $c$ are the gluon polarization and color indices,
respectively. The tree gluon QCD vertex is
$ g_s \bar{\psi}_Q \gamma_\mu \frac{\lambda^c}{2} \psi_Q= g_s
\bar{\psi}_Q \left( \frac{(p_1+p_2)_\mu}{2m_Q} -
\frac{\sigma_{\mu\nu} q_\nu}{2m_Q}\right) \frac{\lambda^c}{2} \psi_Q$.
Comparing its nonrelativistic expansion 
$ \bar{\psi}_Q \gamma_l \psi_Q \simeq (p_1\!+\!p_2)_l\, \varphi_Q^+
\varphi_Q + i \epsilon_{ljk} \, \varphi_Q^+
\sigma_j\varphi_Q \, q_k$ with $T_{ij}(\omega)$ yields $c_G^{\rm tree}=1$. 

To account for the strong interaction corrections we compute
$T_{ij}(\omega)$ perturbatively assuming $-\omega \gg \Lam$. The same
corrections are computed for the $\bar{Q} Q g$ vertex in QCD projected
onto the magnetic spin structure. The vertex yields the one-gluon matrix 
element of the effective heavy quark Lagrangian. 
The difference between the two
determines the coefficient $c_G$. It is ultraviolet (UV) finite, as well as
infrared (IR) finite even at $\vec{q} \to 0$. The latter limit significantly
simplifies computations allowing for directly expanding the Feynman integrands
over $q$. The resulting integrals are saturated in the domain of momenta
between $\omega$ and $m_Q$. The diagrams one has to compute in the
static theory are shown in Figs.~2. 

\begin{figure}[hhh]
 \begin{center} \vspace*{-6mm}
 \mbox{\epsfig{file=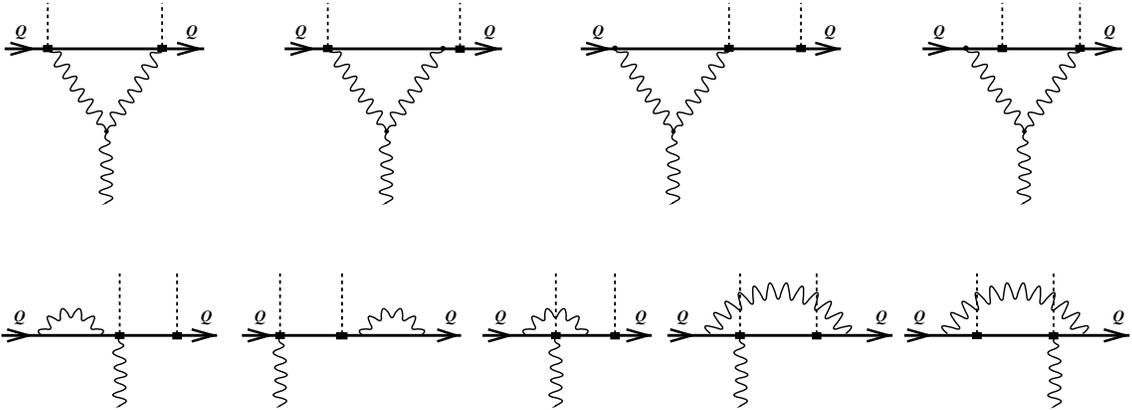,width=15.0cm}} \vspace*{-6mm}
  \end{center}
\caption{
{\small
Examples of one-loop diagrams for the one-gluon matrix element 
of $T_{ij}(\omega)$. The remaining diagrams vanish for the chosen kinematics, 
or upon antisymmetrization over $i,j$. The gluon wavefunction renormalization 
is omitted.}
}\vspace*{-2mm}
\end{figure}

The individual diagrams, however, can be ultraviolet and infrared
divergent. Since the whole integral for $c_G$ is well behaved, one can
use any regularization in the infrared and the ultraviolet for computing
separate diagrams; the only requirement is that it must be consistently
the same. For example, one can compute $c_G$ in dimension
$D\!=\!4\!+\!2\epsilon$; then at $\epsilon \!>\! 0$ there are no IR 
singularity and
the limit $\vec{q} \!\to\! 0$ is straightforward.\footnote{At $D\!=\!4$ 
the chromomagnetic moment in QCD has infrared divergence 
$\sim \ln{1/\vec{q}^{\,2}}$.} On the other hand the point 
$\epsilon\!=\!0$ is regular for $c_G$. 
However, since there are separate diagrams which diverge both in the 
UV and IR, dimensional regularization is not advantageous. Instead, we cut the
integrals in the UV at $k^2\!=\!\Lambda^2$, and introduce the IR mass
regulator in the gluon propagator
$\frac{\delta_{\mu\nu}}{k^2-\lambda^2}$. The latter allows us to use the
limit $\vec{q}^{\,2}\!\to\! 0$ even at $D\!=\!4$. Cancellation of the terms
dependent on $\lambda^2$ and $\Lambda^2$ provides a useful cross-check. 

The computation of the one-loop matrix element of $T_{ij}$ in the
effective static theory results in the Feynman gauge in 
\beq
(T_{ij}\!-\!T_{ji})(\omega) = \frac{1}{-\omega} (\delta_{il} q_j-\delta_{jl} q_i)\,
g_s \bar{Q} \mbox{$\frac{\lambda^c}{2}$} Q\left\{
1 - \frac{g_s^2}{16\pi^2} 
\frac{-C_A}{2}\left( 2\ln{\frac{\Lambda^2}{\omega^2}} - 4
\right)
\right\}
.
\label{22}
\eeq
The QCD vertex takes the form
\beq
\Gamma_\mu= \gamma_\mu - \frac{g_s^2}{16\pi^2} \left\{
\frac{-C_A}{2}\left[ 2\ln{\frac{\Lambda^2}{\lambda^2}}-8 \right]
\gamma_\mu
+ \left[
C_F +
\frac{-C_A}{2}\left(\ln{\frac{m_Q^2}{\lambda^2}} -2\right)
\right] \frac{\gamma_\mu /\!\!\!{q} -\!
/\!\!\!{q}\gamma_\mu}{2m_Q} 
\right\} ,
\label{24}
\eeq
where $C_F=4/3$ and $C_A=N_c$. We have omitted from both expressions the
diagrams renormalizing the external gluon propagator, since they are
the same in both cases. For magnetic structure the difference in the
Abelian part amounts, as expected, to the Schwinger anomalous term
$C_F\frac{\alpha_s}{2\pi}$. The non-Abelian difference is just $-C_A
\frac{\alpha_s}{4\pi} \ln{\frac{m_Q^2}{\omega^2}}$. 

We note, however, that with the non-Abelian interaction the two theories
would have different running strong coupling $g_s(k^2)$ below $m_Q$ if
the bare coupling $g_s^{(0)}$ and the UV cutoff $\Lambda$ are taken the
same. This is seen by evaluating the ``charge" gluon vertex $\Gamma_0$ in
the static theory. The difference originates from the part of the
`Abelian' vertex correction proportional to $C_A$ 
which is not canceled by the 
renormalization of the quark wavefunction -- it yields pure
$\ln{\frac{\Lambda^2}{\lambda^2}}$. (The `non-Abelian' vertex is absent
from the charge interaction.) The total correction to the
$\gamma_\mu$ structure in the vertex in QCD does not depend on the quark mass, 
Eq.~(\ref{24}) as it should be to respect gauge invariance. However,
this holds only provided the UV cutoff is infinitely larger than all
other mass scales. The
renormalization of the static quark interaction differs by a 
constant since here the UV cutoff is much lower than $m_Q$. This means
that the static quark gauge interaction requires a different
counterterm. Alternatively, it can be expressed by saying that for
static quarks $\Lambda_{\rm stat}$ must be taken different from $\Lambda$
in full QCD. As follows from Eq.~(\ref{24}), at one loop one has 
$\Lambda_{\rm stat} \!=\! \Lambda/e^2$ in the $C_A$ term, and 
$\frac{\as}{2\pi}
\,C_A \ln{\frac{m_Q}{\omega}}$ gets replaced by $\frac{\as}{2\pi}\, C_A
\left(\ln{\frac{m_Q}{\omega}}-2\right)$. 

To calculate the matrix element of $\mu_G^2(\mu)$ we need to integrate
$\frac{1}{2\pi}\Im T_{ij}(\omega)$. The literal expression (\ref{22}) is not valid at
$|\omega|\lsim |\vec{q}\,|$ even in perturbation theory. However, the
perturbative $T(\omega)$ has the proper analytic properties to any order.
We then can represent the sum in Eq.~(\ref{10}) as an integral over the
contours in the complex $\omega$ plane stretched away from small
$\omega$, see Fig.~3 where the perturbative matrix element is given by
Eq.~(\ref{22}). The integral is simple:
\beq
\matel{Qg}{g_s G_{ij}(\mu)}{Q} = 
g_s (\delta_{il} q_j-\delta_{jl} q_i)\, \bar{Q} 
\mbox{$\frac{\lambda^c}{2}$} Q
\left\{
1- \frac{g_s^2}{16\pi^2} 
\frac{-C_A}{2}\left( 4\ln{\frac{\Lambda}{e^2 \mu}} - 4
\right)
\right\}\;.
\label{32}
\eeq
We thus get the final result
\beq
c_G(\mu)=
1+ C_F \frac{\alpha_s}{2\pi} + C_A\frac{\alpha_s}{2\pi}
\left(\ln{\frac{\mu}{m_Q}} + 2 \right)\;.
\label{34}
\eeq

\begin{figure}[hhh]
 \begin{center} \vspace*{-7mm}
 \mbox{\epsfig{file=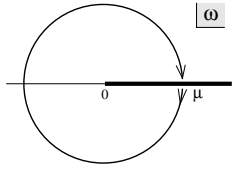,width=3.0cm}} \vspace*{-6mm}
  \end{center}
\caption{
{\small
The complex plane of energy $\omega$. Thick line at $\omega \!>\! 0$
shows the cut of $T(\omega)$. The integral of $\Im T(\omega)$ can
be taken over the circle $|\omega|\!=\!\mu$.}
}\vspace*{-2mm}
\end{figure}

It is convenient to absorb the constant term $2C_A$ in the
non-Abelian part into the argument of the logarithm, {it i.e.}\ use 
$\left(\ln{\frac{\mu}{m_Q}} + 2 \right) =
\ln{\frac{e^{2\!} \mu}{m_Q}}$. Moreover, the usual Abelian part
$\frac{\alpha_s}{2\pi}C_F$ depends on the normalization convention used
for the heavy quark mass. The standard Schwinger coefficient in 
Eq.~(\ref{34}) refers to the
pole mass, a choice disfavored in QCD. Using instead the running
`kinetic' mass
$m_Q(\tilde \mu)$ \cite{fiveblmope,dipole} we find that 
for the chromomagnetic term in the Hamiltonian (\ref{16})
\beq
\frac{c_G(\mu)}{2m_Q}=
\frac{1 + C_A\frac{\alpha_s}{2\pi}\ln{\frac{e^2\mu}{m_Q}}}
{2m_Q(\frac{m_Q}{3})}
\;.
\label{36}
\eeq
Alternatively, using the {\small $\overline{{\rm MS}}$} mass this expression
becomes
\beq
\frac{c_G'(\mu)}{2m_Q}=
\frac{1 + C_A\frac{\alpha_s}{2\pi}\ln{\frac{e^2\mu}{m_Q}}}
{2\,\overline{\!m\!}\,_Q(m_Q/e^{\frac{1}{3}})}
\;;
\label{38}
\eeq
however, the {\small $\overline{{\rm MS}}$} mass of a heavy quark 
loses physical significance at the normalization scales below $m_Q$ 
\cite{rev}.  

\section{Power corrections}

The perturbative relation 
\beq
c_G(\mu) \,\mu_G^2(\mu)= \frac{3}{2}\, m_b\cdot (M_{B^*}-M_B)
\label{39}
\eeq
holds only for asymptotically heavy $b$ where power-suppressed effects 
die out. The $1/m_b^2$ corrections to the hadron masses are given by two
local heavy quark operators and four nonlocal correlators also describing
the $1/m_Q$ corrections to the hadrons' wavefunctions. For
$M_{B^*}\!-\!M_B$ one has \cite{optical}
\beq
M_{B^*}\!-\!M_B = \frac{2}{3} \frac{\mu_G^2}{m_b} + 
\frac{1}{3} \frac{-\rho_{LS}^3}{m_b^2} + 
\frac{1}{3} \frac{\rho_{\pi G}^3+\rho_A^3}{m_b^2}
+ {\cal O}\left(\frac{1}{m_b^3}\right)\;.
\label{40}
\eeq

The expectation values of the {\it convection current\,} (or spin-orbital)
operator $\rho_{LS}^3=
\frac{1}{2M_B}\matel{B}{\vec{\sigma}\!\cdot\! \vec{E}\!\times\!\vec{\pi}}{B}$ 
would vanish to the extent that $B$ could be described as purely a two-body
system like in nonrelativistic approximation. We know, however,  that in
actual QCD the heavy quark bound state is rather relativistic. This is
quantified by the difference between the transition amplitudes
$\tau_{3/2}$ and $\tau_{1/2}$ and between the masses of the $\frac{3}{2}$ and
$\frac{1}{2}$ $P$-wave states, which are the same in nonrelativistic
systems \cite{newsrdurham}. In particular, for the first three moments
we have
\bea
\frac{2\sum_m  |\tau_{3/2}^{(m)}|^2 \!-\! 2\sum_n |\tau_{1/2}^{(n)}|^2}
{2\sum_m |\tau_{3/2}^{(m)}|^2 \,+\, \sum_n |\tau_{1/2}^{(n)}|^2} \;\;\;\;&=&
\frac{1}{2\varrho^2\!-\!\frac{1}{2}} \approx \;0.7\\
\frac{2\sum_m  \vep_m |\tau_{3/2}^{(m)}|^2 \!-\! 
2\sum_n \vep_n|\tau_{1/2}^{(n)}|^2}
{2\sum_m \vep_m |\tau_{3/2}^{(m)}|^2 \,+\, \sum_n \vep_n|\tau_{1/2}^{(n)}|^2} &=&
\;\;\;\frac{2\overline\Sigma}{\La}\;\;\; \approx \;0.7\\
\frac{2\sum_m  \vep_m^2 |\tau_{3/2}^{(m)}|^2 \!-\!
2\sum_n \vep_n^2 |\tau_{1/2}^{(n)}|^2}
{2\sum_m \vep_m^2 |\tau_{3/2}^{(m)}|^2 \,+\, \sum_n
\vep_n^2 |\tau_{1/2}^{(n)}|^2} &=&
\;\;\;\frac{\mu_G^2}{\mu_\pi^2} \;\;\; \approx \;0.8
\eea 
with $\La\simeq 700\MeV$ and $\overline\Sigma \simeq 250\MeV$. The
normalization scale dependent $\varrho^2$, $\La$, $\mu_\pi^2$ and
$\mu_G^2$ are taken at the scale around $1\GeV$.

The magnitude of $\rho_{LS}^3$ can then be estimated using the next spin
sum rule \cite{rev}:
\beq
-\rho_{LS}^3 \simeq \mu_G^2 \, \mu_{\rm hadr} \approx 0.15 \mbox{ to }
0.2\GeV^3\;,
\label{48}
\eeq
where $\mu_{\rm hadr} \approx 500\MeV$ is a characteristic mass scale
for the $P$-wave excitations. The nonlocal correlators are saturated by the transitions into the
$j^P\!=\!\frac{1}{2}^+$ for $\rho_{\pi G}^3$ and into
$j^P\!=\!\frac{1}{2}^+$, $j^P\!=\!\frac{3}{2}^+$ 
for $\rho_A^3$ ``radial'' excitations; they are largely unknown. 

We can estimate the necessary combination of the above spin-triplet
$D\!=\!3$ parameters employing the empirical observation that the mass-square
splitting between vector and pseudoscalar mesons is nearly a constant:
\beq
M_\rho^2-M_\pi^2 \simeq M_{K^*}^2-M_K^2 \simeq M_{D^*}^2-M_D^2 
\simeq M_{B^*}^2-M_B^2
\label{50}
\eeq
which extends even to strange charmed mesons. (A $12\%$ decrease for $B$
fits well the expected perturbative renormalization.) 
It is related to the
universal slope of the corresponding Regge trajectories, a yet poorly
understood nonperturbative phenomenon of strong dynamics,
perhaps related to a certain simplification in the large $N_c$ limit.
This universality must be definitely violated for very heavy quarks due
to hard gluons with momentum scaling with $m_Q$. Rather, it can be
viewed as an inherent property of soft nonperturbative interactions
responsible for physics around $1\GeV$ scale. 

The universality implies the relation
\beq
-(-\rho_{LS}^3+\rho_{\pi G}^3+ \rho_A^3) \simeq 2\La \mu_G^2 
\approx 0.5\GeV^3
\;,
\label{52}
\eeq
where we have used $\La\simeq M_B\!-\!m_b \simeq 700\MeV$. The above
nonperturbative parameters are normalized at the scale around $1\GeV$.
With the estimate (\ref{48}) we are led to an evaluation
\beq
-(\rho_{\pi G}^3+ \rho_A^3)  \approx 0.6 \mbox{ to } 0.7 \GeV^3
\;.
\label{54}
\eeq
The scale of these correlators lies in the expected range if one bears
in mind the magnitude of other parameters $\La$, $\mu_\pi^2$, $\mu_G^2$
all being given by a mass $600\mbox{ to } 700\MeV$ to the
corresponding power, though possibly on the upper side. 

The sign of $\rho_{\pi G}^3$ and $\rho_A^3$ is not known {\it a priori}.
We note, however, that the so far observed nonperturbative phenomena in the
heavy mesons fit well the approximation that the ground state $B$ has 
nearly the ``lowest Landau level'' wavefunction, or is the 
BPS-saturated state. 
For instance, $\mu_\pi^2\!-\!\mu_G^2$ is
noticeably lower than $\mu_G^2$. If this saturation was actually the
case and $\mu_\pi^2\!-\!\mu_G^2=0$, one would necessary have
$\vec\sigma\vec\pi\, \state{B}\!=\!0$, meaning that the asymptotic
wavefunction of the light cloud annihilates two certain linear
combinations of the total momentum operators, e.g.\ ${\cal P}_{\!\!z}$ and
${\cal P}_{\!\!x\,}\!- i{\cal P}_{\!\!y}$ for the state 
$j_z\!=\!\frac{1}{2}$. 
Signifying vanishing of all
$\tau_{1/2}$, this would also entail a series of
relations which include vanishing of all the $B$ meson correlators
involving powers of $\vec\sigma \vec \pi\,$. 
Say, 
\beq 
-\rho_{LS}^3=\rho_D^3\,, \qquad 
\rho_{\pi G}^3 = -2\rho_{\pi\pi}^3\,, \qquad
\rho_A^3 + \rho_{\pi G}^3 = -\rho_{\pi\pi}^3 -\rho_S^3
\label{56}
\eeq
would hold. 
The first relation agrees with the estimate (\ref{48}) (note that
$-\rho_{LS}^3 \ge \rho_D^3$). The correlators $\rho_{\pi\pi}^3$ and
$\rho_S^3$, on the other hand, are positive, which shows consistency of
the estimates. 

Adopting relation (\ref{52}) as the guideline, we end up with
\beq
\mu_G^2 = \frac{3}{4}\: c_G^{-1}(\mu; m_b) \cdot \left(M_{B^*}^2\!-\!M_B^2
\right)
\label{60}
\eeq
through order $1/m_b$. The effect of the $1/m_b$ corrections in
Eq.~(\ref{39}) amounts to the apparent decrease of $\mu_G^2$ by about
$15\%$, recovered in the analysis above. The magnitude of the shift is
moderate and fits well the expectations \cite{vub}. 

\section{Numerical estimates}

Applying equations (\ref{36}), (\ref{40}) and (\ref{52}) we arrive at
the evaluation
\beq
\mu_G^2(2.5\GeV) \simeq 0.312\GeV^2\;,
\label{70}
\eeq
where the value of $m_b(1\GeV)=(4.57\pm 0.05)\GeV$ was used and its
evolution to the scale $1.5\GeV\simeq m_b/3$ can be safely done using
\beq
\frac{{\rm d}m_b(\mu)}{{\rm d}\mu}
=-\left(\frac{16}{3}+\frac{4}{3}\frac{\mu}{m_b}\right) 
\frac{\as^{(d)}(\mu)}{\pi}\;; 
\label{72}
\eeq
the effective dipole radiation strong coupling $\as^{(d)}$ is 
known to two
loops \cite{dipole}. The uncertainty in the mass affects $\mu_G^2$ here
only at a percent level. Likewise, the perturbative effects are included
up to the second loop; their uncertainty can hardly be significant.

The possibly largest uncertainty comes from the precise value of the
$D\!=\!6$ operators. It has been mentioned above that the $12\%$
decrease in $M_{B^*}^2\!-\!M_B^2$ compared to $M_{D^*}^2\!-\!M_D^2$ is
well described by the perturbative renormalization amounting at one loop
to a factor of about $0.87$. However, the two-loop result by Czarnecki
and Grozin \cite{cg} further suppresses $c_G^{(b)}/c_G^{(c)}$ down to
about $0.8$. Taken literally, this would imply even some 
overshooting, so that the power corrections governed by 
$-(\rho_{\!\pi G}^3 \!+\!\rho_{\!A}^3\!-\!\rho_{\!LS}^3)$ have to 
make up for the
extra suppression. This would imply the combination to further exceed
the estimate (\ref{52}) by a factor as large as $1.5\mbox{ to }2$,
being next to fatal for the expansions in $1/m_c$. However, we note that
the major perturbative factor in the extra NLO enhancement of
$M_{D^*}\!-\!M_D$ comes from the increase in the renormalization of the
chromomagnetic moment of the charm quark due to growing $\alpha_s$ at
the charm scale. Introducing the proper Wilsonian cutoff around $1\GeV$
would effectively stop the increase of this short-distance coefficient
for charm, and eliminate the apparent contradiction with the observed
behavior without invoking a too large scale for higher-order power
corrections. 

To remain on the safe side we
allow for an additional factor $0.75\mbox{ to } 1.4$ in Eq.~(\ref{52}). 
Then we arrive at 
\beq
\mu_G^2(2.5\GeV) = (0.30 \mbox{ to } 0.33)\GeV^2
\;.
\label{74}
\eeq

The scale $2.5\GeV$ would be high enough to trust the 
perturbative expansion.
In practical applications we need, however, to evaluate $\mu_G^2$ at the
lower scale around $1\GeV$. The perturbative evolution of $\mu_G^2$ obeys 
\beq
\mu\frac{{\rm d}\mu_G^2(\mu)}{{\rm d}\mu}= 
-C_A \frac{\as^{(me)}(\mu)}{2\pi}\, \mu_G^2(\mu)\;,
\label{76}
\eeq
where $\as^{(me)}$, in analogy with the coupling in Eq.~(\ref{72}) can
be called ``\hspace*{-1.5pt}{\it ME}\,''-radiation one, 
or ``\hspace*{-1.5pt}{\it ME}\,-coupling''. To
first order it is the usual QCD $\alpha_s$.
The complete two-loop calculation presented in paper \cite{cg} combined 
with the one-loop result Eq.~(\ref{34}) is sufficient to determine
$\as^{(me)}(\omega)$ to order $\alpha_s^2$, using renormalization
invariance of observables.  This yields
\beq
C_A \frac{\as^{(me)}(\omega)}{2\pi} = \gamma_m - 2 C_A \left (
\mbox{$\frac{11}{3}$} C_A- \mbox{$\frac{2}{3}$}n_f\right)
\left(\frac{\as}{4\pi}\right)^2 + {\cal O}\left(\as^3 \right)
\;,
\label{77}
\eeq
where $\gamma_m$ is the anomalous dimension of Ref.~\cite{cg}, Eq.~(15)
and $n_f$ is the number of light flavors. Therefore, we have
\beq
\as^{(me)}(\omega) =
\overline{\!\alpha\!}\,_s\left(e ^{-\frac{1}{12}}\, \omega \right)
\left[1-\mbox{$\frac{25}{24}$}\, C_A \frac{\alpha_s}{\pi}\right],
\qquad C_A=N_c=3\;.
\label{79}
\eeq
As expected, the two-loop anomalous dimension of the physically defined
chromomagnetic operator differs from the $\overline{{\rm MS}}$ one
already at order $\alpha_s^2$ (its conformal part is still the same to
two loops). 

The evolution equation Eq.~(\ref{76}) is immediately solved:
\beq
\mu_G^2(\mu) =   \mu_G^2(\omega) 
\left[ 
\frac{\frac{\pi}{\as^{(\!M\!1\!)}(\omega)}+ \frac{\beta_1}{2\beta_0}}
{\frac{\pi}{\as^{(\!M\!1\!)}(\mu)}+ \frac{\beta_1}{2\beta_0}}
\right]^{\frac{C_A}{\beta_0}},
\label{81}
\eeq
where $\beta_0$ and $\beta_1$ are usual QCD one- and two-loop
coefficients in the $\beta$-function, and the evolution of the {\it ME}
coupling to two loops is given by the same renormalization group
equation. 

The evolution equation (\ref{76}) viewed perturbatively suggests that
$\mu_G(\mu)$ is enhanced toward lower normalization scale. Taking it
at face value we get an enhancement factor $1.11$.
This is larger than the literal one-loop difference coming from 
Eq.~(\ref{34}) due to the growing strong coupling which is accounted 
for in the evolution equation. 
One clearly cannot go with the perturbative evolution too low in $\mu$. 
For instance, at $\mu \!\to\! 0$ the value of $\mu_G^2$ goes to
zero, constituting apparently only about $0.2\GeV^2$ at $\mu\!=\!0.5\GeV$
\cite{ioffe,newsrdurham}. 
Nevertheless, it is justified to trust the above moderate enhancement.
The physically defined coupling $\as^{(me)}$ is smaller than
$\overline{\!\alpha\!}\,_s$ and remains suppressed protecting
perturbative corrections from blowing up. This is in contrast with the
standard $\overline{{\rm MS}}$ anomalous dimension which receives large
positive two-loop correction.

Combining the above factors, we arrive at
\beq
\mu_G^2(1\GeV) = \frac{3}{4}\,\left(M^2_{B^*}\!-\!M^2_B \right)\cdot
[\,0.94 \pm 
0.025 {\raisebox{-.5em}{{\tiny\rm pert}}}\,
^{+ 0.06}_{-0.035}
{\raisebox{-.5em}{{\tiny\rm power}}}\,]\;.  
\label{78}
\eeq
Therefore, we conclude that the value of $\mu_G^2(1\GeV)$ most probably
amounts to 
\beq
\mu_G^2(1\GeV) = 0.35 \,^{+ 0.03}_{-0.02}\,\GeV^2\;;
\label{80}                        
\eeq
the lower values can appear in an unfavorable scenario where power 
corrections in charm go out of theoretical control.

\section{Discussion and outlook}

The presented analysis allows us to determine the consistently defined
chromomagnetic value $\mu_G^2$ normalized at the scale around $2\GeV$
with minimal theoretical uncertainty. We conclude that the value
$\mu_G^2(1\GeV)\!\simeq \! 0.4\GeV^2$ we have used so far was
reasonably accurate, yet probably about $10\%$ larger than it is in
reality. In principle, as low a value as $0.30\GeV^2$
cannot be rigorously excluded at present, for the price of endangering
the $1/m_c$ expansions. The uncertainty in the
perturbative effects can be further decreased carrying out the same program to
two loops. A complementary information on the
$D\!=\!3$ nonperturbative parameters in the hadron mass expansion would be
helpful to get more confidence in the evaluation of the power
corrections and to shrink  the error bars down to a few percent level. 

We found the essentially non-Abelian 
effective ``\hspace*{-1.5pt}{\it ME}\,''-coupling 
which does not have a counterpart in electrodynamics, 
either classical or quantum. It  determines, for example, the fraction of 
the total momentum of the QCD degrees of freedom carried by the hard 
(short-distance) components of the light cloud; it is power suppressed, 
$\propto \mu_G^2/\mu^2$  \cite{newsrdurham}. Physics behind this phenomenon 
will be discussed elsewhere.

At the loop level the gluon self-interaction suppresses the effective  
``\hspace*{-1.3pt}{\it ME}\,''-coupling compared to 
the traditional  $\overline{{\rm MS}}$ coupling. The conformal part of it 
turns out similar to the one in the dipole radiation coupling
\cite{dipole}, although the non-Abelian suppression is approximately
twice larger then in the latter, $\frac{25}{24}N_c\,\frac{\alpha_s}{\pi}$ vs.\
$\left(\frac{\pi^2}{6}\!-\!\frac{13}{12}\right)\!N_c\, \frac{\alpha_s}{\pi}$.
The similarity further supports the {\it a priori} expected advantage 
of using the 
radiation coupling as an effective perturbative expansion parameter. 
Following the OPE-based line of reasoning of Ref.~\cite{dipole} we 
find that the $1/\omega$ nonperturbative component in this coupling 
at large $\omega$ is associated with the $LS$ operator and is proportional 
to its anomalous dimension $\gamma_{LS}$.  
Implementing the approach of Ref.~\cite{kinosh} one can show that this 
anomalous dimension exactly coincides with that of the chromomagnetic 
operator.\footnote{The anomalous dimensions of $c_{G\,}\rho_{\pi_G}^3$ and 
$c_{G\,}^2\rho_A^3$ vanish by the same token, and their mixing into 
$\rho_{LS}$ is easily fixed. I am grateful to M.~Eides and A.~Vainshtein 
for discussing this point.}
Then we conclude that in $B$ mesons, for instance, at large $\omega$
\beq
\frac{\delta_{\rm np} \as^{(me)}(\omega)}{ \as^{(me)}(\omega)} 
\simeq   
\frac{-\rho_{\rm LS}^3(\omega)}{\omega \,\mu_G^2(\omega)} \approx 
\frac{0.5\GeV}{\omega}\;.
\label{84}
\eeq
This estimate is consistent with the possibility to
perturbatively evolve $\mu_G^2$ down to a $1\GeV$ scale. 
\vspace*{1mm}

The precise expectation value of the kinetic operator $\mu_\pi^2$ is
quite critical in a number of applications. Heavy quark sum
rules ensure that the inequality $\mu_\pi^2(\mu) > \mu_G^2(\mu)$ holds
at arbitrary normalization point \cite{rev,ioffe}, and $\mu_\pi^2$
almost certainly must lie in the interval $0.4\GeV^2 \le
\mu_\pi^2(1\GeV^2) \le 0.55\GeV^2$. Moreover, using the spin sum rules
\cite{newsrdurham}
one has $\mu_\pi^2(\mu)\!-\!\mu_G^2(\mu) = 3\tilde\vep^2\!\cdot\!
(\varrho^2(\mu)\!-\!0.75)$ with $0.5\GeV \!\lsim \!\tilde\vep\!<\!\mu$,
and we expect $\tilde\vep \!\approx\! 0.5\GeV$ for the usual choice of
$\mu\!=\!1\GeV$.
(One should keep in mind that the
value of the kinetic energy of actual $b$ quark in $B$ meson can be
reduced due to often discarded, but probably noticeable $1/m_b$
effects \cite{vub,lebur}.) 

We anticipate that combining the heavy quark expansion with an 
approximation assuming $\mu_\pi^2\!-\!\mu_G^2 \!\ll\! \mu_G^2$ 
manifesting the proximity
to the ``BPS'' regime for the ground state, can be useful in 
guiding us through understanding pattern, or even physics of higher-order power
corrections in $B$ and $D$ mesons. Expanding around this approximation
provides a new and effective nonperturbative parameter small enough to 
isolate a number of potentially large power corrections. For
example, the usual spin-averaged heavy meson mass expansion replaced by
$$
m_b\!-\!m_c \!=\! M_B\!-\!M_D \!+\! 
\frac{\mu_\pi^2\!-\!\mu_G^2}{2} 
\left(\!\frac{1}{m_c}\!-\!\frac{1}{m_b}\! \right) \!-\! 
\frac{-\!\rho_D^3\!-\!\rho_{LS}^3 \!+\! 
\rho_{\pi\pi}^3 \!+\!\rho_{\pi G}^3  \!+\!\rho_{s}^3 \!+\!\rho_{A}^3}{4}
\left(\!\frac{1}{m_c^2}\!-\!\frac{1}{m_b^2} \!\right)
$$
\vspace*{-4mm}
\beq
\qquad \qquad \qquad \qquad + \;{\cal O}\left(\frac{1}{m_Q^3}\right)
\label{90}
\eeq
is possibly more stable in respect to higher orders being governed by
the suppressed higher-dimension expectation values scaling like powers of 
$\sqrt{\mu_\pi^2 \!-\!\mu_G^2}\,$. As has been pointed out \cite{vadem},
the uncertainty in the precise value of $m_b\!-\!m_c$ is currently the main
limiting factor in extracting $|V_{cb}|$. 
The validity of the approximation can
be experimentally cross checked by carefully analyzing the semileptonic
$b\to c$ transitions into excited states, in particular to the
$\frac{1}{2}$ $P$-waves.

In the BPS regime, a number of other nonperturbative effects become more
tractable. Say,
the $B\!\to \!D$ formfactor at zero recoil does not have $1/m_Q^2$
corrections at all. The corrections still are present and significant 
in the $B\!\to\! D^*$ zero recoil formfactor $F(0)\,$, but their structure
simplifies:
\beq
F(0) = \xi_A^{\frac{1}{2}} - (1\!+\!\chi)\,\frac{\mu_G^2}{6m_c^2} -
{\cal O}\left(\frac{1}{m_c^3}\right)
\label{92}
\eeq
with $\chi \,\mu_G^2$ given by a sum of two other positive nonlocal
correlators $\tilde \rho_{\pi\pi}^2 + \tilde \rho_{S}^2$ 
{\small
$$
\chi\,\mu_G^2 = \!\int \!i|x_0|\,{\rm d}^4x\, \frac{1}{4M_{\!B}}\!
\left[\matel{B}{iT\{\bar{b}\vec{\pi}^2b(x), \bar{b}\vec{\pi}^2
b(0)}{B}'\!
+ \frac{1}{3} 
\matel{B}{iT\{\bar{b}\sigma_j B_k b(x), \bar{b}\sigma_j B_k b(0)}{B}'
\right]\!,
$$
} 
\hspace*{-.45em}similar to $\rho_{\pi\pi}^3$ and $\rho_{S}^3$ 
(see Eqs.~(28) of Ref.~\cite{optical}), but containing the extra factor of
$i|x_0|$ in the integrand, or $1/(E_n\!-\!E_0)$ in the language of 
perturbation theory in quantum mechanics. 
We expect the approximation $\tilde \rho_{\pi\pi}^2 \!+\! \tilde \rho_{S}^2
\approx (\rho_{\pi\pi}^3 \!+\! \rho_{S}^3)/\vep_{\rm rd}$ (actually, a
rigorous upper bound \cite{optical}) to be reasonably accurate, 
with $\vep_{\rm rd}\!\approx\! 600\MeV$ 
the energy of the first $j^P\!\!=\!\frac{1}{2}^+$ radial excitation of
the ground state. Taken literally this would suggest $\chi$ to be quite 
large, around $2$.

The limit $\mu_\pi^2\!-\!\mu_G^2 \!=\! 0$ means that the heavy quark
wavefunction minimizes the momentum square operator in a given
chromomagnetic field. This happens for the lowest Landau level which is
an example of a ``BPS-saturated'' state. It is worth noting that the newer
relativistic quark models of heavy hadrons \cite{orsayqm} properly
implementing Lorentz transformations yield a good 
approximation to this limit. Complementary to this, the Block \& Shifman 
QCD sum rule analysis of the IW function \cite{bsrho} strongly
supports this by virtue of the spin sum rules.
In usual quantum mechanical
systems of electrons the BPS saturation is realized applying 
strong magnetic field.\footnote{It is interesting
to recall that the limit of a strong magnetic field in a quantum
mechanical system of charged particles yields a physical realization of
noncommuting space coordinates \cite{jackiw}. Actual $B$ mesons 
may well share some of their properties.}
In mesons the chromomagnetic field is {\it a priori} of the same order
$\Lam^2$ as the chromoelectric field, and is far from classical. In
$B$ mesons the approximate BPS saturation would rather be dictated by a very
special internal structure of the light cloud, leading to the strong
correlation between the spin and momentum operator. Such a correlation
vanishes in a nonrelativistic system, and can be realized only in a
deeply relativistic regime. It does not look probable that such a
property, if confirmed experimentally, is purely accidental.  Perhaps, it
is related to a certain large parameter, like the number of space
dimensions or the number of colors. It is intriguing to study such 
possible connections.
\vspace*{1.5mm}

Our analysis of $\mu_G^2$ incorporating powerful constraints from 
a series of the heavy quark sum rules gives an indication that 
the higher-dimension
expectation values are quite significant and probably lie at the higher
end of the existing estimates. Likewise, the pattern of the excited
heavy quark states seems to be at some variance with the routinely
employed assumptions inherited from obsolete models. There are fresh
ideas of how to improve the knowledge in less vulnerable ways. These
will be addressed in forthcoming publications.
\vspace*{.3cm}\\
{\bf Acknowledgments:} ~The author is pleased to thank M.~Eides for
illuminating conversations and I.~Bigi for useful discussion of possible
applications; help by A.~Czarnecki, P.~Nason and M.~Shifman is gratefully
acknowledged. I am thankful to A.~Vainshtein for both criticism 
and constructive suggestions. 
This work was supported in part by the NSF under grant number PHY00-87419.

\end{document}